\documentclass[a4paper,12pt]{article}

\usepackage{amsmath,amssymb,latexsym,amsfonts,amsthm}
\usepackage{enumerate}
\usepackage[longnamesfirst]{natbib}

\theoremstyle{plain}
\newtheorem{Thm}{Theorem}[section]

\theoremstyle{definition}

\newtheorem{Def}[Thm]{Definition}
\newtheorem{Rem}[Thm]{Remark}

\newcommand{\bE}{\ensuremath{\mathbf{E}}}
\newcommand{\bP}{\ensuremath{\mathbf{P}}}
\newcommand{\bR}{\ensuremath{\mathbf{R}}}

\numberwithin{equation}{section} 

\title{Semi-Static Hedging Based on a Generalized 
Reflection Principle on a Multi Dimensional 
Brownian Motion}
\author{
Yuri Imamura\thanks{ 
Department of Mathematical Sciences, Ritsumeikan University. 
Email: {\tt yuri.imamura@gmail.com}} 
and Katsuya Takagi\thanks{Department of Mathematical Sciences, Ritsumeikan University. 
}}
\date{}

\begin{document}

\maketitle

\begin{abstract}
On a multi-assets Black-Scholes 
economy, we introduce a 
class of barrier options, where the knock-out boundary 
is a cone.   
In this model we apply a generalized reflection principle in 
a context of 
the finite reflection group acting on a Euclidean space to give a 
valuation formula and 
 the semi-static hedge. 
The result is a multi-dimensional 
generalization of the {\em put-call symmetry} 
by \citet{BC}, \citet{CC}, etc. 
The important implication of our result is 
that with a given volatility matrix structure 
of the multi-assets, one can design a multi-barrier option
and a system of plain options, with the latter the former is 
statically hedged.  
\end{abstract}
\section{Introduction}
In this paper, we introduce a 
class of barrier options (knocked-in and 
knocked-out options) on a multi-assets Black-Scholes 
economy and give a semi-static hedging 
technique based on a generalized reflection principle in 
a context of 
the finite reflection group acting on a Euclidean space. 
A semi-static hedging strategy 
will be obtained from an equation between 
the value of barrier options 
and that of path-independent options. 

\subsection{The previous studies on one risky asset cases}
Before going into details, we give a short survey of the context. 
\citet{Merton} was the first to discuss
a hedging strategy of a barrier option. It was 
in the Black-Scholes framework. 
The valuation formula was in closed form but using a dynamic 
delta hedging strategy in the underlying asset. 
\citet{BC} gave a hedging strategy of barrier options 
using {\it put-call symmetry} in the Black-Scholes framework. 
Their strategy is 
semi-static in that it is a portfolio of a few 
European options with a fixed maturity 
rebalanced at most one time.

The put-call symmetry is a geometric generalization of 
the reflection principle, which is roughly state as the following
relation on the process $ S $; 
\begin{equation}\label{PCSp}
\bE_0 [ f (S_T) 1_{\{ \tau_H > T \}}] = 
\bE_0 [1_{\{S_T > H \}} f (S_T) ] 
- \bE_0 [1_{\{S_T < H \}} f (H^2 /S_T) (S/H)^p ]
\end{equation}
for any bounded $ f $, where 
$ \tau_H $ is the first hitting time of the process to the 
boundary $ H > 0 $.  The power $ p $ is related to the {\em drift};
$ p =1 $ when $ S $ is a martingale. One sees why it is called 
{\em put-call symmetry} by looking at 
the case $ f (x) = (x -H)_+ $. 
Detailed discussions, applications, and extensions are found 
in \citet{BC}, \citet{CEG}, \citet{Poulsen}, \citet{CC} and \cite{CL}.

\subsection{The multi-risky asset case}
As mentioned above, in this paper we are interested in the static hedge of 
knock-out options written on {\bf multi-assets}. 
There have been several attempts to work on barrier options of  
multi-assets (\citet{MS}, \citet{Schmutz}, etc) but they are basically 
working on one reflection; the group structure is trivial. 
The present paper in contrast heavily relies on the group structure. 

Suppose that the prices of the assets are given by 
strictly positive adapted processes 
$ S^i $, $ i= 1, \cdots, n $, defined on a filtered probability space 
$ (\Omega, \mathcal{F}, P, \{ \mathcal{F}_t \} ) $. We assume that $ 0 $-th 
asset is non-risky; $ S^0 $
is a strictly positive deterministic process and that 
$ S^i/S^0 $, $ i=1,2, \cdots, n $ are strictly positive martingales.
By the fundamental theorem 
of asset pricing, it may mean that the market is arbitrage-free and complete
and $ P $ is already the unique equivalent martingale measure.  

The pay-off of a knock-out option in its generic form is 
\begin{equation}\label{KO1}
f(S_T)1_{ \{ \tau \geq T \}},
\end{equation}
where $f:\mathbf{R}^{n+1} \rightarrow \mathbf{R}$ is 
a measurable function 
and $ \tau_D $ is the first exit time 
of the $ n +1 $-dimensional process $ S = (S^0, S^1, \cdots, S^n ) $ 
out of a region $ D \subset \mathbf{R}^{n+1} $, which should be a
stopping time with respect to the market filtration $ \{\mathcal{F}_t \} $.  
Following \cite{Imamura}, we formulate the possibility of a static hedge 
as follows. 
\begin{Def}
Let $ g $ 
be a measurable function on $ \mathbf{R}^{n+1} $. 
We say a knock-out option (\ref{KO1}) is 
{\bf statically hedged}
by the path-independent option (whose pay-off is) $ g $
if it holds 
\begin{equation}\label{SHF}
\bE_0 [ 1_{ \{ \tau \geq T \}} f(S_T) | \mathcal{F}_t ]
= 1_{ \{ \tau \geq t \}} 
\bE_0 [ g (S_T) | \mathcal{F}_t ], \quad t \in [0,T], 
\end{equation}
assuming that both sides are finite. 
\end{Def}

Financially speaking, the left-hand-side (\ref{SHF}) 
stands for the price at time $ t $ of the knock-put option, 
while the right-hand-side is the value of the path-independent 
option until the exit time $ \tau $. The equality claims that 
the former is hedged by holding the latter 
until $ \tau $. At the time $ \tau $, 
both sides becomes zero, meaning in particular that 
the value of the path-independent 
option can be sold without any cost. 

The main result of the present paper is 
Theorem \ref{Theorem-ref}, where we will establish 
\begin{equation}\label{SHIT}
\begin{split}
& \bE_0[1_{ \{\tau_{D_H} > T \}} f (S_T^C) |\mathcal{F}_t ] \\
& = 
1_{ \{\tau_{D_H} > t \}}
\bE_0 \left[ \sum_{ w \in W (\Phi)}\varepsilon(w) f (HS_T^{C_w} /H^{T_w} ) 
S_T^{ ^t x_w C}/H^{^t x_w} 1_{ \{ S_T^{C_w} > H^{T_w} \}}
| \mathcal{F}_t \right].
\end{split}
\end{equation}
Here, $ S $ is a given multi-dimensional geometric Brownian motion, 
and $ H $ is a given (multi-) boundary, and $ W (\Phi) $ is the reflection 
group; 
detailed descriptions to them 
will be given in section \ref{GRP1} and section \ref{Section3-reflection}. 
The implication of (\ref{SHIT}) is that, given 
$ S $, one can choose $ C $ so that we can utilize the group structure of
$ W (\Phi) $ to construct a static hedge of the option. 

In Remark \ref{REd}, 
we show that Theorem \ref{Theorem-ref} is actually a 
generalization of the put-call symmetry (\ref{PCSp}). 

\subsection{The organization of the present paper}

In section \ref{GRP1} we first recall the Brownian motion 
in a fundamental domain of the 
action of a finite reflection group. 
Various descriptions of $n$-dimensional Brownian motion 
conditioned for its components to never collide and more 
generally of Brownian motion 
in the fundamental domain have been studied extensively in recent time, 
by for example \citet{Gra}, \citet{OC}, \citet{Bia}, \citet{JoOC} 
and references therein. 
Such a reflected Brownian motion is a generalization of the
reflecting Brownian motion in $[0,\infty)$. 

In Section \ref{Section3-reflection}, 
we give the main result (Theorem \ref{Theorem-ref}) of the present paper. 
Using the generalized
reflection principle on the Brownian motion, we give 
semi-static hedging strategies of  knock-out options in 
a multi-stock model where the knock-out boundary is related to 
a root system. 

\cite{Akahori-Takagi} is the earliest related paper, where 
only a simplest case is studied. 
\cite{Akahori-Imamura} is the latest one, where {\em symmetrization} 
of diffusion processes,  
a technique for numerical analysis of barrier options, is introduced, 
where a generalization of the reflection principle, 
which is still based on the symmetry of reflection groups, 
plays a central role.

\section{Generalized reflection principle}\label{GRP1}
\subsection{The finite reflection groups}\label{sec1}
We recall basics of the finite reflection groups from \citet{Hall}. 
Let $V$ be a real finite dimensional Euclidean space endowed with an 
inner product $\langle u,v \rangle$ for $u,\, v \in V$. 
A reflection is a 
linear operator $s_{\alpha}$ on $V$ which 
transforms a nonzero element $\alpha \in V$ to 
the one reflected with respect to the hyperplane 
$H_{\alpha}:=\{x \in V:\ \langle x, \alpha \rangle=0\}$ 
orthogonal to $\alpha$. 
The reflection is explicitly given by: 
\begin{equation*}
s_{\alpha}(x)
=  x - 
\frac{2\langle x, \alpha\rangle}{\langle \alpha, \alpha \rangle}\alpha\ 
(x \in V). 
\end{equation*}

Since we attach greater importance to an element $\alpha$ of $V$ than 
the hyperplane $H_{\alpha}$, we call $s_{\alpha}$ {\it a reflection 
operator with respect to} $\alpha$. It is easy to see that 
$s_{\alpha}$ is an orthogonal transformation. Since $s_{\alpha}$ move 
$\alpha$ to $-\alpha$, it holds that $s_{\alpha}^2=1$. It means 
$s_{\alpha}$ has order $2$ in the group $O(d)$ of 
all orthogonal transformations of $V$. 
Take $\Phi$ to be a finite set such that nonzero 
vectors in $\Phi$ 
satisfying the two conditions:
\begin{itemize}
\item[(R1)] 
$\Phi \cap \bR \alpha = \{\alpha, -\alpha\} $ for all $\alpha \in \Phi$
\item[(R2)] 
$s_{\alpha} \Phi = \Phi$ for any $\alpha \in \Phi$.
\end{itemize}
Then a group $W(\Phi)$ generated by reflections 
$\{s_{\alpha}:\ \alpha \in \Phi\}$
(we call it a {\it reflection group} in short) 
is a finite subgroup of $O(d)$. 
We call $\Phi$ a {\it root system} associated with a reflection group 
because of the connection between 
Weyl groups and root systems of semi-simple Lie algebras. 

Our definition of root systems differs somewhat from that commonly used in 
Lie theory. A finite reflection group is characterized by our root system, and 
conversely, the group generated by a root system is finite. 

Fix a root system $\Phi$. 
A {\it fundamental system} is a subset $\Sigma$ 
of a root system $\Phi$ that satisfies the following properties: 
\begin{itemize}
\item[(F1)]
The elements of $\Sigma$ are linearly independent. 
\item[(F2)]
Every element of $\Phi$ can be written as a linear combination of the 
elements of $\Sigma$ with coefficients all of the same sign 
(all non-positive or all nonnegative). 
\end{itemize}
Then the reflection group $W(\Phi)$ is a group generated by 
reflections $s_{\alpha}$ for $\alpha \in \Sigma$, and subject 
only to the relations: $( s_{\alpha}s_{\beta})^{m(\alpha, \beta)}=1$ for 
$\alpha,\ \beta \in \Sigma$. (See, for example, Section $1.9$ in \citet{H}.)
Because of (R$1$), $\{w(\alpha):\ w \in W(\Phi),\ \alpha \in \Sigma\}$ 
is equal to the root system $\Phi$, therefore the root system 
is retrieved by the associated fundamental system. 

For each $\alpha \in \Sigma$, we set the open half-space 
$A_{\alpha}:=\{x \in V:\ \langle x, \alpha \rangle >0 \}$. As 
an intersection of open convex sets, 
$C_{\Sigma}:= \cap_{\alpha \in \Sigma} A_{\alpha}$ is an open convex set. 
We call $C_{\Sigma}$ a {\it chamber} 
corresponding to a fundamental system $\Sigma$. 
Let $\bar{C}_{\Sigma}$ to be the closure of $C_{\Sigma}$. 
Note that the hyperplane $\cup_{\alpha \in \Phi}H_{\alpha}$ is the 
boundary $\partial C_{\Sigma}:= \bar{C}_{\Sigma} \setminus C_{\Sigma}$. 
We note that a chamber 
$C_{\Sigma}$ is a interior set of a fundamental domain 
for the action of $W(\Phi)$ on $V$, that is, each 
$x \in V\setminus \cup_{\alpha \in \Phi} H_{\alpha}$ is conjugate under $W(\Phi)$ to one and only one element in $C_{\Sigma}$. 
It is clear that replacing $\Sigma$ by $\omega \Sigma$ replaces $C_{\Sigma}$ by $\omega C_{\Sigma}$ for $\omega \in W(\Phi)$. 
Thus the chambers are characterized topologically as the 
connected components of the complement 
in $V$ of $\cup_{\alpha \in \Phi} H_{\alpha}$. 
Conversely, elements of a root system can be characterized as 
vectors which are orthogonal to some wall of a chamber. 
\subsection{Brownian motion in a chamber 
corresponding to a fundamental system of a root system}
Let $B=(B_t)_{t\geq 0}$ be a standard Brownian motion in $\bR^d $
and $\bP_x$ be a Wiener measure with the initial distribution $\delta_x$. 
The transition density of Brownian motion in $\bR^d$ is given by 
\begin{equation*}
p_t(x,y)=(2 \pi t)^{-d/2} \exp\left(-\frac{| y-x |^2}{2t}
\right),\ x,y\in \bR^d \ {\rm and}\ t \geq 0. 
\end{equation*}
Note that $ p_t $ is invariant under the action of 
the group $O(d)$ of 
all orthogonal transformations of $\bR^d$ and
hence a reflection group. 

Fix a root system $\Phi$ and its fundamental system $\Sigma$. 
For a chamber $C_{\Sigma}$, 
let $T_{C_{\Sigma}}:= \inf \{t \geq 0 :\ B_t \in 
\hspace{-1em}/ 
\ C_\Sigma\}$ 
be the first hitting time on the boundary 
$\partial C_{\Sigma}$. 
We denote by $\hat{p}_t(x,dy)$ the transition density of Brownian motion 
in $C_{\Sigma}$ killed at the boundary $\partial C_{\Sigma}$, that is, 
\begin{equation*}
\hat{p}_t(x,A)= \bP_x ( B_t \in A,\ T_{C_{\Sigma}}>t )
\end{equation*}
for any Borel set 
$A$ of $\bar{C_{\Sigma}}$. We can reconstruct $\hat{p}_t$ by using $p_t$ 
and a generalized reflection principle (see for details \citet{MR0114248}) as: 
\begin{equation}\label{GRP}
\begin{split}
\hat{p}_t(x,dy) &= \sum_{w \in W(\Phi)} \varepsilon(w) p_t(x,w(y))dy\\
\end{split}
\end{equation}
for $x,y \in C_{\Sigma}$, where 
\begin{eqnarray*}
\varepsilon (w)= \det w =
\left\{ \begin{array}{ll}
1 & {\rm if }\ 
w=s_{\alpha_1}s_{\alpha_2}\cdots s_{\alpha_k}, 
\ {\rm and}\ k\ {\rm is\ even}.  \\
-1 & {\rm if }\ 
w=s_{\alpha_1}s_{\alpha_2}\cdots s_{\alpha_k}, 
\ {\rm and}\ k\ {\rm is\ odd}. \\
\end{array} \right.
\end{eqnarray*} 

In the following sections, 
we aim to apply this expression to a static hedging strategy of 
a knock-out option in the context of mathematical finance. 

\section{The Static Hedge of Options Knocked Out at the 
Boundary of a Chamber Corresponding to 
a Fundamental System of a Root System}\label{Section3-reflection}
\subsection{Knock-out Option on a multi-asset Black-Scholes type model 
}\label{reflection-section2}
We work on the following Black-Scholes economy: 
the price processes $ S_t^i $, $ i =0,1,\cdots,n $ 
are given by 
\begin{equation}\label{model00}
{S^i_t} = S_0^i \exp \left\{ ( \Lambda B_t)_i  + (r - \frac{1}{2} 
(\Lambda \Lambda^*)_{i,i} ) t \right\}, 
\,\ S^i_0 >0, 
\end{equation}
where 
$ B $ is a standard $d$-dimensional ($ d \leq n $) Brownian motions 
starting from $ 0 $ 
$ \Lambda $ is an $ (n +1) \times d $ matrix with 
$ \mathrm{rank} \Lambda = d $, and  
$r$ is the risk-free interest rate. Here $ 0 $-th row vector of 
$ \Lambda $ is set to be zero so that 
we have 
\begin{equation*}
S^0_t= {S^0_0}\exp ( r t). 
\end{equation*}

Let $\Lambda_l$ and $\Lambda_r$ be a $d \times (n+1)$ matrix 
and a $d \times d$ matrix such that $\Lambda_l \Lambda = E_d$ and 
$\Lambda \Lambda_r = E_r$, respectively. Since 
$ \mathrm{rank} \Lambda = d $, $\Lambda_l$ and $\Lambda_r$ are exist. 

Let $ m $ be an integers smaller than $ d $, and 
$ C = (c_{k,i}) $ be
an $ m \times (n+1) $ matrix with $ \mathrm{rank}\, C=m $. Define 
\begin{equation*}
D_H := \bigcap_{k=1}^m 
\left\{ s = (s^0,s^1, \cdots, s^n) 
\in \mathbf{R}^{n+1} : \prod_{i=0}^n (s^i)^{c_{k,i}} \leq H_k \right\},
\end{equation*}
where $ H = (H_1, \cdots, H_m ) \in \mathbf{R}^m_{++} $.
This is not an empty set by the assumption that $ \mathrm{rank} \, C=m $. 
We study the possibility of the static hedge 
of the knock-out options whose pay-off is 
$ f (S_T) 1_{ \{ \tau_D \geq T \} } $. Recall that 
$ \tau_{D_H} $ is the first exit time of $ S $ out of $ D_H $. 
To see if $ P (\tau_{D_H} < T) > 0 $ or not, 
we rewrite all the settings in terms of the Brownian motion $ B $.
Define 
\begin{equation}\label{G}
G := 
\left\{ x = (x^1, \cdots, x^d) 
\in \mathbf{R}^{d} : C \Lambda x  \in \mathbf{R}^m_+ \right\}.
\end{equation}
Note that $ G  $ is a convex cone $ \ne \emptyset $
since $ \mathrm{rank} \, C \Lambda = m $. 

Let $ \mu $ and $ h $ be such that 
\begin{equation}\label{def-mu}
C \Lambda \mu = C \{ r \mathbf{1} - \frac{1}{2} 
\mathrm{diag} (\Lambda \Lambda^*)\} 
\end{equation}
and
\begin{equation}\label{def-h}
C \Lambda h = C \log S_0 - \log H,  
\end{equation}
where $ \mathbf{1} $ denotes the vacuum vector $ (1, 1, \cdots,1) $,
and $ \mathrm{diag} : M (n+1) \to \mathbf{R}^{n+1} $ sends
a matrix to the vector composed of its diagonal entries. 
Such $ \mu $ and $ h $ exist since $ \mathrm{rank} \, ( C \Lambda ) = m $. 

With these, we have that 
\begin{equation*}
\tau_{D_H} = \inf \{ t>0: h+B_t + \mu t \not\in G \}. 
\end{equation*}
Now we see that $ P (0<\tau_{D_H} <T ) > 0 $. 

\subsection{Possibility of Static Hedge}\label{MT}
Before stating our main result, we 
introduce some notations. 
For vectors $ a $ and $ b $, 
map $ \phi: \bR \rightarrow \bR $ and 
$ k \times l $ matrix
$ M = (m_{i,j}) $, 
we mean 
$ (a_1 b_1, \cdots, a_k b_k)$ by 
$ a b $,  
$ (a_1/b_1, \cdots, a_k/b_k)$ by 
$ a/b $, 
$(\phi (a_1), \cdots, \phi (a_k))$ 
by $\phi (a) $, 
$$ ( \prod_{j=1}^k v_j^{m_{1,j}}, \cdots, \prod_{j=1}^k 
v_j^{m_{l,j}}) $$ 
by $ \mathbf{v}^M $,
and $a < b$ means $a_j < b_j$ for all $1 \leq j \leq k$. 

Now we are in a position to state our main result. 
Let $ \Sigma =\{\alpha_1 , \alpha_2 \cdots, \alpha_m\}$ 
be a fundamental system of a root system $\Phi$ on $\bR^{d}$ with 
$ \sharp \Sigma = m $. 
Let $ C $ satisfy $ C \Lambda = \Sigma' $, 
where $i$-th row of the $ m \times d$ 
matrix $\Sigma'$ is $ \nu_i \alpha_i $, for some 
$ (\nu_1, \cdots, \nu_m ) \in \bR^{m}_{++}$. 
Since $ \Lambda $ and $ \Sigma' $ is of full rank (the latter is implied by
(F1) in section \ref{sec1}),
such $ C $ with $ \mathrm{rank}\, C = m $ exists. 
Note that $ \mathrm{span} \Sigma = C \Lambda $ is 
invariant under the action of $ \omega \in W (\Phi) $, 
which is implied by (F2) in section \ref{sec1}.  
The representation matrix for the action by $ w $ will be denoted by
$ T_w \in GL(m,\mathbf{R}) $. That is, 
\begin{equation}\label{repmat}
 C \Lambda w = T_w C \Lambda.
\end{equation} 

With the choice of $ C $ subject to the volatility $ \Lambda $ and 
a fundamental system $ \Phi $, we have the following 
\begin{Thm}\label{Theorem-ref}
(i) The set 
$ G $ given by (\ref{G})  
is a fundamental domain of the action of the reflection 
group $ W (\Phi) $. 
(ii) In such a case, any measurable $ f : \mathbf{R}^{m}_{++}  
\to \mathbf{R} $ with at most linear growth,
$ f (S_T^C) 1_{ \{\tau_{D_H} > T \}} $ is statically hedged by
\begin{equation*}
\sum_{ w \in W (\Phi)}\varepsilon(w) f (HS_T^{C_w} /H^{T_w} ) 
S_T^{ ^t x_w C}/H^{^t x_w} 1_{ \{ S_T^{C_w} > H^{T_w} \}}, 
\end{equation*}
where $ C_w = T_w C $ and 
$ x_w \in \mathbf{R}^m $ is a solution to $ ^t \Sigma' x_w = w (\mu) - \mu $,
which surely exists by the linearly independence among the roots. 
\end{Thm}
\begin{proof}
By the Markov property, it suffices to show 
\begin{equation*}
\bE_0[f (S_T^C) 1_{ \{\tau_{D_H} > T \}}] = 
\bE_0[\sum_{ w \in W (\Phi)}\varepsilon(w) f (HS_T^{C_w} /H^{T_w} ) 
S_T^{ ^t x_w C}/H^{^t x_w} 1_{ \{ S_T^{C_w} > H^{T_w} \}}].
\end{equation*} 
Applying the Cameron-Martin-Maruyama-Girsanov theorem, 
we obtain that 
\begin{eqnarray}\label{proofeq-1}
\begin{split}
&\bE_0[f (S_T^C) 1_{ \{\tau_{D_H} > T \}} ]\\
&=\bE_0[f (S_0^C 
e^{C\{\Lambda (B_T - \mu t ) +(r\mathbf{1} - \frac{\mathrm{diag} \Lambda \Lambda^*}{2})T\}
}) 
\\
& \hspace{3cm}
e^{\langle \mu, B_T\rangle  - \frac{|\mu|^2 T}{2}}:\inf\{t > 0 : B_t + h \not\in G \} > T ],\\
\end{split}
\end{eqnarray}
where $\langle \cdot, \cdot \rangle$ is the Euclidean inner product and 
$|\cdot |$ is the Euclidean norm. 
By 
shifting the initial point of Brownian motion to $h$, 
\begin{equation}\label{proofeq-2}
\begin{split}
&\mbox{(the right hand side of (\ref{proofeq-1}))}\\
&=\bE_h[
f (S_0^C 
e^{C\{\Lambda (B_T- h - \mu T) +( r\mathbf{1} - \frac{\mathrm{diag} \Lambda \Lambda^*}{2})T\}} ) 
\\
& \hspace{3cm}
e^{\langle \mu, B_T - h \rangle  - \frac{|\mu|^2 T}{2}}:\inf\{t > 0 : B_t \not\in G \} > T ].\\
\end{split}
\end{equation}
By the expression (\ref{GRP}), we have 
\begin{equation}\label{proofeq-3}
\begin{split}
&\mbox{(the right hand side of (\ref{proofeq-2}))}\\
&=\bE_h[\sum_{\omega \in W(\Phi )} \varepsilon(w) 
f (S_0^C 
e^{C\{\Lambda (w(B_T)- h - \mu T) +( r\mathbf{1} - \frac{\mathrm{diag} \Lambda \Lambda^*}{2})T\}} ) 
\\
& \hspace{3cm}
e^{\langle \mu, w(B_T) - h \rangle  - \frac{|\mu|^2 T}{2}}
: w(B_T) \in G  ]. 
\end{split}
\end{equation}
Since $w$ is an orthogonal transformation, 
$\langle \mu, w(B_T)\rangle$ is equal to $ \langle w^{-1}(\mu), B_T\rangle$. 
We note that $w^{-1} = w$ since it is a reflection; has the order $2$. 
Therefore we see that 
\begin{equation}\label{proofeq-4}
\begin{split}
&\mbox{(the right hand side of (\ref{proofeq-3}))}\\
&=\bE_h[\sum_{\omega \in W(\Phi )} \varepsilon(w) 
f (S_0^C 
e^{C\{\Lambda w B_T - \Lambda ( h + \mu T) 
+( r\mathbf{1} - \frac{\mathrm{diag} \Lambda \Lambda^*}{2})T\}} ) 
\\
& \hspace{3cm}
e^{\langle w(\mu), B_T \rangle - \langle \mu, h \rangle - \frac{|\mu|^2 T}{2}}
: w(B_T) \in G  ].
\end{split}
\end{equation}
By 
shifting the initial point again, we obtain that 
\begin{equation}\label{proofeq-5}
\begin{split}
&\mbox{(the right hand side of (\ref{proofeq-4}))}\\
&=\bE_0[\sum_{\omega \in W(\Phi )} \varepsilon(w) 
f (S_0^C 
e^{C\{\Lambda w (B_T + h ) - \Lambda ( h + \mu T)
+(r\mathbf{1} - \frac{\mathrm{diag} \Lambda \Lambda^*}{2})T\}} ) 
\\
& \hspace{3cm}
e^{\langle w(\mu), B_T  \rangle 
+ \langle w(\mu) - \mu, h \rangle - \frac{|\mu|^2 T}{2}}
: w(B_T + h) \in G  ] \\
&=\bE_0[\sum_{\omega \in W(\Phi )} \varepsilon(w) 
f (S_0^C 
e^{C\{\Lambda w (B_T + h ) - \Lambda ( h + \mu T) 
+( r\mathbf{1} - \frac{\mathrm{diag} \Lambda \Lambda^*}{2})T\}}) \\
& \hspace{3cm}
e^{\langle \mu, B_T\rangle  - \frac{|\mu|^2 T}{2}}
e^{\langle w(\mu) - \mu, B_T + h \rangle}
: w(B_T + h) \in G  ].
\end{split}
\end{equation}
Using the Cameron-Martin theorem again, we see that 
\begin{equation}\label{proof-eq-last}
\begin{split}
&\mbox{(the right hand side of (\ref{proofeq-5}))} \\
&=\bE_0[\sum_{\omega \in W(\Phi )} \varepsilon(w) 
f (S_0^C 
e^{C\{\Lambda w (B_T +\mu T + h ) - \Lambda ( h + \mu T) 
+( r\mathbf{1} - \frac{\mathrm{diag} \Lambda \Lambda^*}{2})T\}} ) 
\\
& \hspace{3cm}
e^{\langle w(\mu) - \mu, B_T +\mu T  + h \rangle  }
: w(B_T +\mu T + h) \in G  ]\\
&=\bE_0[\sum_{\omega \in W(\Phi )} \varepsilon(w) 
f (
He^{C\{\Lambda w (B_T +\mu T + h ) \} } ) 
\\
& \hspace{3cm}
e^{\langle w(\mu) - \mu, B_T +\mu T  + h \rangle  }
: w(B_T +\mu T + h) \in G  ],
\end{split}
\end{equation}
where the last equality of (\ref{proof-eq-last}) comes from 
(\ref{def-mu}) and (\ref{def-h}).

Therefore, we have to show 
\begin{equation}\label{insidef}
HS_T^{C_w}/H^{T_w} 
=e^{C\{\Lambda w (B_T +\mu T + h ) + \log H \}} ,
\end{equation}
\begin{equation}\label{density}
S_T^{ ^t x_w C}/H^{ ^t x_w } 
=
e^{\langle w(\mu) - \mu, B_T +\mu T  + h \rangle  },
\end{equation}
and
\begin{equation}\label{domains}
\{S_T^{C_w} > H^{T_w}\}
=\{w(B_T +\mu T + h) \in G \}.
\end{equation}
The relation (\ref{insidef}) is equivalent to 
\begin{eqnarray}
C_w \Lambda &=& C \Lambda w, \label{cw1} \\
 C_w \log S_0+ \log H -\log H^{T_w} &=& C \Lambda w h + \log H, \label{cw2}
\end{eqnarray}
and 
\begin{equation}
C_w ( r\mathbf{1} - \frac{\mathrm{diag} \Lambda \Lambda^*}{2})  = 
 C \Lambda w \mu. \label{cw3}
\end{equation}
By the defining relation (\ref{repmat}) of $C_w$, 
(\ref{cw1}) is valid. 
Substituting (\ref{def-mu}), (\ref{def-h}) and (\ref{cw1}), 
we also confirm (\ref{cw2}) and (\ref{cw3}).
Also, the relation (\ref{density}) is equivalent to 
\begin{eqnarray}
 ^t x_w C \Lambda &=&w(\mu) - \mu \label{tw1}, \\
 ^t x_w (C \log S_0 -\log H)  &=& (w(\mu) - \mu) h \label{tw2},
\end{eqnarray}
and 
\begin{equation}
 ^t x_w C ( r\mathbf{1} - \frac{\mathrm{diag} \Lambda \Lambda^*}{2})  
=  (w(\mu) - \mu) \mu. \label{tw3}
\end{equation}
Similarly as the case of (\ref{insidef}), 
(\ref{tw1}), (\ref{tw2}) and (\ref{tw3}) 
are valid.
Finally, the right hand side of (\ref{domains}) is equal to 
\begin{equation*}\label{domain}
\{C \Lambda w(B_T +\mu T + h)\in  \bR_+^m \}
= \{e^{C \Lambda w(B_T +\mu T + h)} > \mathbf{1}\}.\\
\end{equation*}
Hence (\ref{domains}) is satisfied due to (\ref{insidef}). 
\end{proof}
\begin{Rem}\label{REd}
We show that 
in the case of $1$-dimension, 
Theorem \ref{Theorem-ref}
reduces to the put-call symmetry (\ref{PCSp}) by 
\citet{BC}. 
Suppose that in the Black-Scholes economy; 
there is a risky asset with volatility $\sigma$, 
and the interest rate is $r$. 
We take a fundamental system $\Sigma = \{1\}$ on $\bR$. 
Then the root system generated by $\Sigma$ is 
$\Phi= \{1,-1\}$, and the reflection group is 
$W(\Phi) = \{1, s_1\}$, 
where we recall that $s_1 (x) = -x$ for $x \in \bR$. 
Then we obtain that $T_1 x = x$ and $T_{s_1} x=-x$  for $x \in \bR$.
We take $\nu = 1$. Then $C = 1$, 
$\mu = r - \frac{1}{2} \sigma^2$, 
$x_1 = 0$ and $x_{s_1}= 1 - \frac{2r}{\sigma^2}$. 
For $H >0$, we set 
$D_H = \{ s \in \bR: s \leq H\}$. 
By Theorem \ref{Theorem-ref}, 
$f(S_T)I_{\tau_{D_H} > T} $ 
can be hedged by the following: 
\begin{equation}
\begin{split}
f(S_T) I_{\{S_T >H\}} - 
f(\frac{H^2}{S_T})(\frac{S_T}{H})^{ 1 - \frac{2r}{\sigma^2}} I_{\{S_T <H\}}.
\end{split}
\end{equation}
Now we notice that $ p $ in (\ref{PCSp}) is 
$ 1 - \frac{2r}{\sigma^2} $. 
\end{Rem}
\subsubsection*{Acknowledgments}
The authors are grateful to Professors Jir\^o Akahori,  
for many valuable comments, careful reading the
manuscript and suggesting several improvements. 
The authors wish to express their thanks to 
anonymous referee.

\end{document}